\documentclass[showpacs,aps,prl,twocolumn,superscriptaddress]{revtex4-1}

\usepackage[latin1]{inputenc}
\usepackage{graphicx}
\usepackage{hyperref}
\usepackage{epstopdf}
\usepackage{dcolumn}
\usepackage{bm}
\usepackage{amsmath}
\usepackage{amssymb}
\usepackage[dvipsnames]{xcolor}
\usepackage{color}
\usepackage{ulem}
\usepackage{xspace}
\usepackage{ulem}

\usepackage{wasysym}
\usepackage{float}

\renewcommand{\vec}[1]{\boldsymbol{#1}}

\newcommand{\eg}[0]{e.g.\@\xspace}

\newcommand{\las}[0]{\langle}
\newcommand{\ras}[0]{\rangle}
\newcommand{\llas}[0]{\langle\langle}
\newcommand{\rras}[0]{\rangle\rangle}

\renewcommand{\tilde}[1]{\widetilde{#1}}

\newcommand{\ve}[1]{\boldsymbol{#1}}

\usepackage[extra]{tipa}

\setcitestyle{square,numbers,compress}
\usepackage{here}

\begin{document}


\title{Simulation of Fermionic and Bosonic Critical Points with Emergent SO(5) Symmetry}

\author{Toshihiro Sato}
\affiliation{\mbox{Institut f\"ur Theoretische Physik und Astrophysik, Universit\"at W\"urzburg, 97074 W\"urzburg, Germany}}
\author{Zhenjiu Wang}
\affiliation{Max-Planck-Institut f\"ur Physik komplexer Systeme, Dresden 01187, Germany}
\author{Yuhai Liu}
\affiliation{\mbox{School of Science, Beijing University of Posts and Telecommunications, Beijing 100876, China}}
\author{Disha Hou}
\affiliation{\mbox{Department of Physics, Beijing Normal University, Beijing 100875, China}}
\author{Martin Hohenadler}
\affiliation{\mbox{Institut f\"ur Theoretische Physik und Astrophysik, Universit\"at W\"urzburg, 97074 W\"urzburg, Germany}}
\affiliation{Independent Researcher, Josef-Retzer-Str.~7, 81241 Munich, Germany}
\author{Wenan Guo}
\affiliation{\mbox{Department of Physics, Beijing Normal University, Beijing 100875, China}}
\affiliation{\mbox{Beijing Computational Science Research Center, 10 East Xibeiwang Road, Beijing 100193, China}}
\author{Fakher F. Assaad}
\affiliation{\mbox{Institut f\"ur Theoretische Physik und Astrophysik, Universit\"at W\"urzburg, 97074 W\"urzburg, Germany}}
\affiliation{\mbox{W\"urzburg-Dresden Cluster of Excellence ct.qmat, Am Hubland, 97074 W\"urzburg, Germany}}

\begin{abstract}
We introduce a model  of  Dirac  fermions in  2+1 dimensions with a semimetallic,
a quantum spin-Hall insulating (QSHI), and an s-wave superconducting (SSC) phase. The
phase diagram features a multicritical point at which all three phases meet as
well as a QSHI-SSC deconfined critical point. The QSHI and SSC orders
correspond to   mutually  anti-commuting mass  terms of the Dirac Hamiltonian.
Based  on this  algebraic property,  SO(5)  symmetric  field  theories
have been  put forward  to  describe both types of critical points. Using
quantum Monte Carlo simulations, we directly study the operator that
rotates between QSHI and SSC states. The results suggest that it commutes with
the low-energy effective Hamiltonian at criticality but has a gap in the ordered
phases. This implies an emergent SO(5) symmetry at both the multicritical and the deconfined
critical points.
\end{abstract}

\maketitle

{\it Introduction.}---Within the renormalization group theory of phase
transitions \cite{Goldenfeld}, criticality is defined by scale invariance and operators are
classified as either relevant, marginal, or  irrelevant. The concept of an
emergent symmetry refers to critical points that have a higher symmetry than the
corresponding ultraviolet (UV) model as a result of the irrelevance of operators
that break said symmetry. For instance, in interacting one-dimensional (1D)
systems described by 1+1D field theories, emergent Lorentz symmetry is  the rule
\cite{Giamarchi} and leads to the interchangeability of space and time. In 2+1D, 
O($N$) non-linear  sigma  models  are  robust  only for
$N < 3$  \cite{Carmona00,Calabrese03} so  that O(2) symmetry  can emerge in a Z$_4$
 invariant model  \cite{Shao20}. In
other cases, emergent symmetries allow to rotate one  ordered state into
another. For example, the SO(5) theory of high-temperature
superconductivity \cite{ZhangSC97}  conjectured  the unification  of the d-wave superconducting
and  antiferromagnetic orders.   Away  from a critical point with emergent
symmetry, the operator describing the above rotation acquires a gap and is
expected to manifest itself as a resonance with specific quantum numbers
in spectroscopy measurements.

\begin{figure}[t]
  \centering \includegraphics[width=0.35\textwidth]{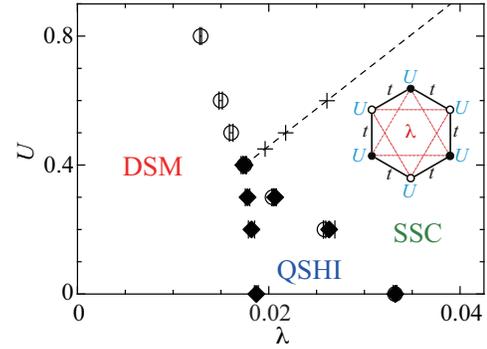}
  \caption{\label{fig:phasediagram} Ground-state phase diagram with Dirac
    semimetal (DSM), quantum spin-Hall insulating (QSHI), and s-wave
    superconducting (SSC) phases from QMC simulations (see text).
    The dashed line and crosses indicate the values of $g$ (0, 0.05, 0.1, 0.2) used in 
    Fig.~\ref{fig:CSO5}(b). Inset: honeycomb plaquette illustrating the
    hopping and interaction terms in Eq.~(\ref{eq:HlambdaU}).  }
\end{figure}

Dirac systems are a fruitful platform to investigate emergent
symmetries \cite{Janssen18}.  Let us consider the 2+1D case of four
two-component Dirac fields relevant for graphene \cite{Neto_rev}. In this setting, one
can define quintuplets of mutually anti-commuting mass terms of either two
Kekul\'e and three antiferromagnetic masses or two s-wave superconducting (SSC)
and three quantum spin-Hall insulator (QSHI) masses, respectively \cite{Ryu09}.
Each quintuplet can be associated with a 5D superspin order parameter
\cite{Tanaka05}.  Theories in which Dirac fermions couple symmetrically to the
superspin have SO(5) symmetry \cite{WangZ20}. However, UV models of interacting Dirac fermions
generically do not. A key question is, therefore, if the symmetry
emerges at critical points, of which we consider two classes. First,
Gross-Neveu fermionic critical points \cite{Herbut09} at which the superspin vector
vanishes and which separate a Dirac semimetal (DSM) from an ordered phase (\eg, the
DSM-QSHI and DSM-SSC transitions in Fig.~\ref{fig:phasediagram}). For this
case, results from an $\epsilon$-expansion \cite{Janssen18} predict an emergent
SO(5) symmetry. The second class are bosonic critical points, where amplitude fluctuations of
the superspin can be neglected and the gapped fermions can be integrated out
\cite{Abanov00}. This case is described by an SO(5) symmetric non-linear sigma
model with a Wess-Zumino-Witten geometrical term, which has been suggested to
describe deconfined quantum critical points (DQCPs) separating two phases with
different order parameters \cite{Senthil04_2,Senthil06,WangZ20}.

In this Letter, we introduce and simulate a suitable lattice model. It has a
global SU(2)$\times$U(1) symmetry associated with spin rotation symmetry and
charge conservation, respectively. The model permits
auxiliary-field quantum Monte Carlo (QMC) simulations without a sign problem and
supports DSM, QSHI, and SSC phases, see Fig.~\ref{fig:phasediagram}. Based on
measurements of the dynamics of the operator that rotates between QSHI and SSC
states, we argue that SO(5) symmetry indeed emerges both at a DSM-QSHI-SSC
Gross-Neveu multicritical point and at QSHI-SSC DQCPs, at least at the
intermediate energy scales accessible in our simulations. 

{\it Model and Method.}---We consider the Hamiltonian
\begin{eqnarray}
\label{eq:HlambdaU}
\hat{H}&=&
  -t\sum_{\langle \ve{i}, \ve{j} \rangle}\left(\hat{\boldsymbol{c}}^{\dag}_{\ve{i}} \hat{\boldsymbol{c}}^{}_{\ve{j}}+\text{H.c.}\right)
  -\lambda\sum_{\hexagon} \Bigg( \sum_{ \langle \langle \ve{i},\ve{j} \rangle \rangle  \in \hexagon }\hat{J}_{\bm{i},\bm{j}} \Bigg)^2 \nonumber \\
   && -U\sum_{\ve{i}} \left(\hat{n}_{\ve{i}\uparrow}-\frac{1}{2} \right) \left(\hat{n}_{\ve{i}\downarrow}-\frac{1}{2} \right)\,.
\end{eqnarray}
Here, $\hat{c}_{\ve{i}\sigma}^{\dagger} (\hat{c}_{\ve{i}\sigma} )$ creates
(annihilates) a fermion with spin $\sigma = \uparrow, \downarrow$ at site
$\ve{i}$ of a honeycomb lattice. The first term in Eq.~(\ref{eq:HlambdaU}) describes nearest-neighbor
hopping of amplitude $t$. The second term defines an interaction on a hexagonal plaquette
between next-nearest-neighbor pairs of sites $\llas \ve{i},\ve{j}
\rras$ (see inset of Fig.~\ref{fig:phasediagram}) with $ \hat{J}_{\bm{i},\bm{j}}
= i \nu_{ \bm{i} \bm{j} } \hat{\ve{c}}^{\dagger}_{\bm{i}} \bm{\sigma}
\hat{\ve{c}}^{\phantom\dagger}_{\bm{j}} + \text{H.c.}$, $\hat{\boldsymbol{c}}^{\dag}_{\ve{i}} =(\hat{c}^{\dag}_{\ve{i},\uparrow},\hat{c}^{\dag}_{\ve{i},\downarrow}
)$, the Pauli vector $\boldsymbol{\sigma}=(\sigma^x,\sigma^y,\sigma^z)$, and phase factors
$\nu_{\boldsymbol{ij}}=\pm1$ as in the Kane-Mele model \cite{KaneMele05b,Liu18}.
The last term is an attractive, onsite Hubbard interaction ($U>0$) with
$\hat{n}_{\ve{i}\sigma} \equiv \hat{c}_{\ve{i},\sigma}^\dagger \hat{c}^{\phantom{\dagger}}_{\ve{i},\sigma}$,
see Fig.~\ref{fig:phasediagram}. In addition to the global SU(2)$\times$U(1)
symmetry discussed above, $\hat{H}$ is  invariant under a
particle-hole transformation so that our choice of chemical potential $\mu=0$
corresponds to half-filling ($\las \hat{n}_{\ve{i}\sigma}\ras=1/2$).

Hamiltonian~(\ref{eq:HlambdaU}) was simulated using the ALF (Algorithms for
Lattice Fermions) implementation \cite{ALF_v1,ALF_v2} of the grand-canonical, finite-temperature,
auxiliary-field QMC method~\cite{Blankenbecler81,Assaad08_rev}.
A sign problem is absent since, after a Hubbard-Stratonovitch transformation
to decouple the perfect square interaction terms, time-reversal
symmetry  as  well as  U(1)  charge  conservation are present for each field configuration
\cite{Wu04}. Results were obtained on lattices with $L\times L$ unit cells
($2L^2$ sites) and periodic boundary conditions. We use units where
$\hbar=k_\text{B}=t=1$ and set $\Delta\tau=0.2$ for the Trotter discretization.

\begin{figure}[t]
  \centering \includegraphics[width=0.48\textwidth]{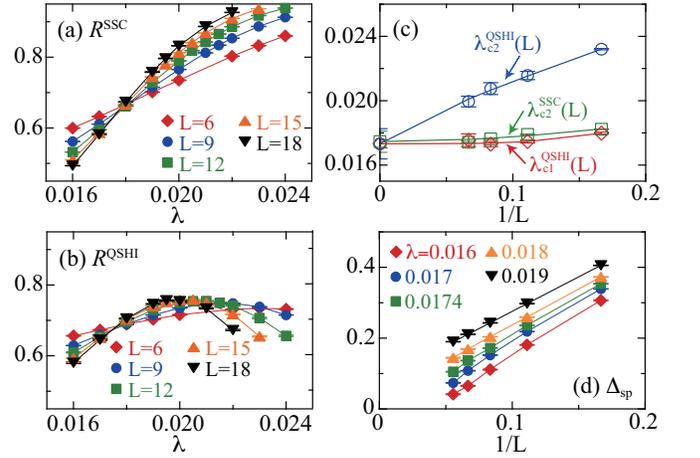}
  \caption{\label{fig:Rchai} Correlation ratios for (a) SSC order and (b) QSHI
    order at $U=0.4$ for different $L$ and $\beta=L$. Extrapolation of the
    crossing points of results for $L$ and $L+3$ in (c) yields a single critical
    point at $\lambda_c\approx0.0174$.  (d) Single-particle gap at the Dirac
    point $\ve{q}= \ve{K}$ near $\lambda_c$ at $U=0.4$. Here, $T=1/30$,
    representative of the ground state for the parameters
    shown.}
\end{figure}

\textit{ Phase diagram.}---To determine the phase boundaries, we computed
the susceptibilities of operators of spin-orbit coupling,
$\hat{\boldsymbol{O}}^{\text{QSHI}}_{\boldsymbol{r},\boldsymbol{\delta}} =
\mathrm{i}\hat{\boldsymbol{c}}^{\dagger}_{\boldsymbol{r}} \boldsymbol{\sigma}
\hat{\boldsymbol{c}}^{}_{\boldsymbol{r}+ \boldsymbol{\delta}} + \text{H.c.}$, and
on-site s-wave paring,
$\hat{O}^\text{SSC}_{\ve{r},\ve{\tilde{\delta}}} =\frac{1}{2}
\hat{c}^{\dagger}_{\ve{r} +\ve{\tilde{\delta}},\uparrow}
\hat{c}^{\dagger}_{\ve{r} +\ve{\tilde{\delta}},\downarrow} + \text{H.c.}$
Here, $\ve{r}$ specifies a unit cell containing  $A$ and $B$ orbitals as well as a hexagon,
$\ve{r} + \ve{\delta}$ runs over hexagon, and $\ve{r} +\ve{\tilde{\delta}}$ over the two
orbitals of the unit cell.  The susceptibilities read
\begin{eqnarray}
  \label{eq:chi}
  \chi^{\alpha}_{\ve{\delta},\ve{\delta}'} (\boldsymbol{q})
  =\frac{1}{L^2} \sum_{\boldsymbol{r},\boldsymbol{r'}}\int_{0}^{\beta} \text{d} \tau
  e^{\mathrm{i}\boldsymbol{q}\cdot(\boldsymbol{r}-\boldsymbol{r}')}
  \langle
  \hat{\boldsymbol{O}}^\alpha_{\boldsymbol{r},\boldsymbol{\delta}}(\tau)
  \hat{\boldsymbol{O}}^\alpha_{\boldsymbol{r'},\boldsymbol{\delta}'}(0)
  \rangle \nonumber
\end{eqnarray}
with $\beta=1/T$ and $\alpha=\text{QSHI}$, {SSC}.  After diagonalizing the matrices
$\chi^{\alpha}_{\ve{\delta},\ve{\delta}'}(\boldsymbol{q})$, we
extracted the renormalization-group invariant correlation ratios \cite{Binder1981,Pujari16}
\begin{eqnarray}
  R^{\alpha}=1-\frac{\chi^{\alpha}({\ve q}_0+\delta {\ve q})}{\chi^{\alpha}({\ve q}_0)}.
  \label{eq:CR-SM}
\end{eqnarray}
Here, $\chi^{\alpha}({\ve q})$ is the largest eigenvalue of $\chi^{\alpha}_{\ve{\delta},\ve{\delta}'} (\boldsymbol{q})$,
${\ve q}_0$ the ordering wave vector, and ${\ve q}_0 + \delta {\ve q}$ the
longest wave-length fluctuation of the ordered state for a given lattice size.
Long-range order in channel $\alpha$ implies a divergent corresponding
susceptibility $\chi^{\alpha} \equiv \chi^{\alpha}({\ve q_0=0})$. Accordingly,
$R^{\alpha} \to 1$ for $L\to\infty$,
whereas $R^{\alpha} \to 0$ in the disordered phase.  At the critical
point, $R^{\alpha}$ becomes scale-invariant for sufficiently large system size $L$, leading
to a crossing of results for different $L$. We assumed a dynamical
critical exponent $z=1$ to set $L=\beta$ in the finite-size scaling, as
justified by the emergent Lorentz invariance of the corresponding field
theory \cite{Herbut09,Senthil04_2,Gross74}.

Figure~\ref{fig:phasediagram} shows the resulting ground-state phase diagram
in the  $\lambda$--$U$ plane. Previous work at $U=0$ revealed consecutive
DSM-QSHI and QSHI-SSC quantum phase transitions with increasing $\lambda$
\cite{Liu18,Liu21}. In particular, the DSM is
stable up to a nonzero critical interaction due to its vanishing density of
states \cite{Sorella92}. The DSM-QSHI transition is an example of a
Gross-Neveu critical point \cite{Herbut09}, whereas the QSHI-SSC transition can be understood in
the framework of DQCPs \cite{Senthil04_1}. We find that the additional Hubbard
interaction favors SSC order, reducing the extent of the QSHI phase with
increasing $U$. For $U\gtrsim 0.5$, we observe a direct DSM-SSC transition with
increasing $\lambda$ that is expected to be in the previously studied U(1) Gross-Neveu universality
class~\cite{Li17,Otsuka18} (see the SM \cite{SM}). A key feature of the phase
diagram of Hamiltonian~(\ref{eq:HlambdaU}) is the existence (within our accuracy)
of a multicritical point at which the DSM, QSHI, and SSC phases meet.

Detailed results for $U=0.4$ are presented in Fig.~\ref{fig:Rchai}. The data for
$R^\text{SSC}$ in Fig.~\ref{fig:Rchai}(a) are consistent with a transition to
the SSC phase at $\lambda_{c2}^{\text{SSC}}=0.0174(6)$, the value obtained by extrapolating
the crossing points to $L\to\infty$ in Fig.~\ref{fig:Rchai}(c). The analysis of
$R^\text{QSHI}$ at $U=0.4$ is more involved. Figure~\ref{fig:Rchai}(b) suggests
two phase transitions and an intermediate QSHI phase, as observed for smaller
$U$. However, the extrapolation of the two sets of crossing points, shown in
Fig.~\ref{fig:Rchai}(c), reveals a single transition at the same value, 
$\lambda_{c1}^\text{QSHI}=0.0174(2)$ and $\lambda_{c2}^\text{QSHI}=0.0174(8)$,   and hence a multicritical point at
$(\lambda_c,U_c)\approx(0.0174, 0.4)$. At this point, according to
Fig.~\ref{fig:Rchai}(d), the single-particle gap vanishes.  The latter was
extracted from the asymptotic behavior of the imaginary-time
single-particle Green function~\cite{Assaad96a}.  Evidence for a continuous
transition in terms of the free-energy derivative as well as results for other
values of $U$ can be found in \cite{SM}.

{\it Multicritical point.}---We now turn to the nature of the
multicritical point. A possible field-theory description is  based on a
16-component  spinor $\hat{\Psi}^{\dagger}_{\nu,\mu,\tau,\sigma}(\ve{k}) $ with Bogoliubov ($\nu$),
valley ($\mu$),  orbital ($\tau$),  and spin ($\sigma$) indices.  Specifically,
 $ \hat{\Psi}^{\dagger}_{1,\mu,\tau,\sigma}(\ve{k})  = \hat{c}^{\dagger}_{\tau, \mu \ve{K} + \ve{k}, \sigma} $ and
 $ \hat{\Psi}^{\dagger}_{-1,\mu,\tau,\sigma}(\ve{k})  = \hat{c}^{}_{\tau, \mu \ve{K} - \ve{k}, \sigma} $,  with
the Dirac points $\pm \ve{K} $.   In this  basis,  the  Dirac  Hamiltonian reads
\begin{equation}
  \hat{H}_0    =     -\frac{v_\text{F}}{2}   \sum_{\ve{k}}  \hat{\ve{\Psi}}^{\dagger} (\ve{k})
  (   k_x  \tau^{x} \mu^{z}- k_y \nu^{z} \tau^y )
  \hat{\ve{\Psi}}(\ve{k})\,,
\end{equation}
 where the Pauli matrix $\tau^{\alpha}$ acts on $\tau$ and likewise for the
 other indices.  The ordered phases observed numerically
 correspond to five mutually anti-commuting mass terms $\hat{M}_i$.  For
 instance, the three QSHI mass terms read
 $\hat{\ve{M}}^\text{QSHI} = \sum_{k} \hat{\ve{\Psi}}^{\dagger} ( \nu^z
   \sigma^{x},\sigma^{y},\nu^{z}\sigma^{z}) \mu^z\tau^z \hat{\ve{\Psi}}$
 and the two SSC masses are given by
 $\hat{\ve{M}}^\text{SSC} = \sum_{k} \hat{\ve{\Psi}}^{\dagger} ( \nu^y,
   \nu^x) \sigma^y\mu^x \hat{\ve{\Psi}}$. The Gross-Neveu Lagrangian
 expected to describe the multicritical point is
\begin{equation}
  \label{eq:GN}
  {\cal L}   =   {\cal L}_0   +  g  \ve{\varphi}(\ve{x}) \cdot  \ve{M}(\ve{x})   +  {\cal L}_\text{B}( \ve{\varphi} )\,,
\end{equation}
where $ \ve{\varphi}(\ve{x}) $ is a five-component field at a point $\ve{x}$ in
2+1D Euclidean space, ${\cal L}_0$ is the Lagrangian density of the free
Dirac system and ${\cal L}_\text{B}( \ve{\varphi} ) $ that of the bosonic field.
Both, ${\cal L}_0$ and $g \ve{\varphi}(\ve{x}) \cdot \ve{M}(\ve{x}) $ are invariant under SO(5)
rotations generated by $\frac{i}{2} [ \hat{M}_i, \hat{M}_j] $.
However, $ {\cal L}_\text{B}( \ve{\varphi} ) $ is only invariant under the
SO(3)$\times$SO(2) rotation of the order parameter vector.
%

\begin{figure}[b]
  \centering \includegraphics[width=0.48\textwidth]{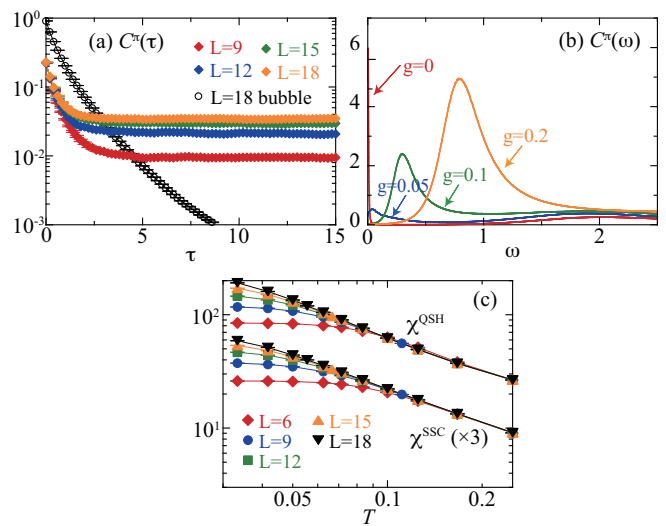}
  \caption{\label{fig:CSO5} (a) Time-displaced correlation function of
    $\hat{\pi}$, $C^{\pi} (\tau)$, at the multicritical point ($U=0.4$,
    $\lambda=0.0174$) for $T=1/30$. Also shown is the result of a bubble
    calculation.  (b) Dynamical structure factor $C^{\pi} (\omega)$ along the dashed
    line in Fig.~\ref{fig:phasediagram}.
    Here, $T=1/30$ and $L=18$.  (c) Temperature dependence of the QSHI and SSC
    susceptibilities at $U=0.4$, $\lambda=0.0174$.  }
\end{figure}

Based on an $\epsilon$-expansion, it is argued in Ref.~\cite{Janssen18} that the
terms that reduce the symmetry from  SO(5) to SO(3)$\times$SO(2) are irrelevant
at the multicritical point. To obtain numerical  evidence, we
consider  the generator  of  SO(5)  that rotates between QSHI and SSC order,
given by $\frac{i}{2}[\hat{M}^\text{QSHI}_z,   \hat{M}^\text{SSC}_1 ] $. A
lattice realization of this operator takes the form
$\hat{\pi}=\sum_{\ve{r}}\hat{\pi}_{\ve{r}}$ with \cite{SM}
\begin{eqnarray}
\label{Eq:pi-mode_main}
  \hat{\pi}_{\ve{r}}  &=&
  \hat{\ve{a}}^{\dagger}_{\ve{r}} \sigma^x  \hat{\ve{a}}^{\dagger}_{\ve{r} + \ve{a}_1}
  +  \hat{\ve{a}}^{\dagger}_{\ve{r}+ \ve{a}_1} \sigma^x  \hat{\ve{a}}^{\dagger}_{\ve{r} + \ve{a}_2}
  +  \hat{\ve{a}}^{\dagger}_{\ve{r}+ \ve{a}_2} \sigma^x  \hat{\ve{a}}^{\dagger}_{\ve{r} }  \\ \nonumber
  && -\, \hat{\ve{b}}^{\dagger}_{\ve{r}} \sigma^x  \hat{\ve{b}}^{\dagger}_{\ve{r} + \ve{a}_1}
  -  \hat{\ve{b}}^{\dagger}_{\ve{r}+ \ve{a}_1} \sigma^x  \hat{\ve{b}}^{\dagger}_{\ve{r} + \ve{a}_2}
  -  \hat{\ve{b}}^{\dagger}_{\ve{r}+ \ve{a}_2} \sigma^x
      \hat{\ve{b}}^{\dagger}_{\ve{r} }   + \text{H.c.}
\end{eqnarray}
Here, $\hat{\vec{a}}^{\dagger}_{\ve{r}}  =  (  \hat{a}^{\dagger}_{\ve{r},\uparrow}, \hat{a}^{\dagger}_{\ve{r},\downarrow} )$ and
$\hat{a}^{\dagger}_{\ve{r},\sigma}$ creates an electron in orbital $A$ of unit
cell $\ve{r}$; a similar definition holds for $ \hat{\ve{b}}^{\dagger}_{\ve{r}} $.
The operator $\hat{\pi}$ transforms as the $z$-component of a vector under spin 
rotations. It is odd under inversion and time reversal \cite{SM} and breaks the
U(1) charge symmetry. Our use of the same notation as in the SO(5) theory of
high-temperature superconductivity \cite{ZhangSC97} is motivated by an
expected resonance in neutron scattering experiments at the
antiferromagnetic wave vector $(\pi,\pi)$ (being odd under inversion) inside the
SSC phase with broken U(1) symmetry.

In Fig.~\ref{fig:CSO5}(a), we plot QMC results for
$C^{\pi} (\tau)= \langle \hat{\pi}(\tau)\hat{\pi} (0)\rangle $ at
$(\lambda,U)=(0.0174, 0.4)$, the estimated location of the multicritical
point. The fact that $C^{\pi} (\tau)$ is independent of $\tau$ at large $\tau$
has important implications. The time $\tau_1$ beyond which $C^{\pi}
(\tau)\approx\text{const.}$ defines an energy scale $\Lambda = \frac{1}{\tau_1}$
as well as a projection onto a low-energy Hilbert space,
$ \hat{P} = \sum_{E_n - E_0 < \Lambda } | n \rangle \langle n | $ with
$\hat{H}| n \rangle = E_n | n \rangle $.  In the latter, the
$\tau$-independence of $C^{\pi} (\tau)$ is equivalent to the statement that
$ \hat{P} [ \hat{H}, \hat{\pi} ] \hat{P} = 0 $ \cite{SM}.  Precisely
the same holds for a conserved quantity such as the total charge at the UV
scale. In this case, $\tau_1$ vanishes and the Hamiltonian commutes with the
total particle number. From these arguments and the data in
Fig.~\ref{fig:CSO5}(a), we infer that $\hat{\pi}$ commutes with the low-energy
effective Hamiltonian, $\hat{H}_\text{eff} = \hat{P} \hat{H} \hat{P}$. This in
turn implies an emergent SO(5) symmetry.

Figure~\ref{fig:CSO5}(a) also includes results obtained by neglecting vertex
contributions. This \textit{bubble} approximation to $C^{\pi} (\tau)$ exhibits a
clear decay even at large $\tau$, revealing that our findings are linked to
interactions in the particle-particle channel.

Because the $\pi$ mode carries charge and spin, it is expected to acquire an
energy gap both in the QSHI and the spin-singlet SSC phases.  This can be
verified by considering the corresponding dynamical structure factor,
$C^{\pi}(\omega) = \text{Im} \chi(\omega) / \left( 1 - e^{-\beta \omega} \right)
$ with
$\chi(\omega)=i \int_0^{\infty} dt \, e^{i\omega t} \left< \left[ \hat{\pi},
    \hat{\pi}(-t) \right] \right>$.  We computed this quantity using 
stochastic analytical continuation \cite{Beach04a}, as implemented in the ALF
\cite{ALF_v2} library. The results in Fig.~\ref{fig:CSO5}(b) are for different
distances $g$ from the multicritical point along the path shown in
Fig.~\ref{fig:phasediagram}. They confirm that the $\pi$
mode is gapless at criticality ($g=0$) but has a gap that increases with $g$ as
we go deeper into the SSC phase.

Further evidence for an emergent SO(5) symmetry can be obtained from the
temperature dependence of the QSHI and SSC susceptibilities.  At the
multicritical point, and given Lorentz invariance, they are expected
to scale as $\chi^{\alpha} \sim L^{(2-\eta^{\alpha})}f(\beta/L)$, with anomalous
dimensions $\eta^{\alpha}$ \cite{Vicari14}. This is borne out by the QMC results in
Fig.~\ref{fig:CSO5}(c), which exhibit similar behavior at low temperatures and
identical values of $\eta$ within error bars, namely
$\eta^{\text{QSHI}}=0.99(3)$ and $\eta^{\text{SSC}}=1.01(1)$.

\begin{figure}[t]
  \centering \includegraphics[width=0.48\textwidth]{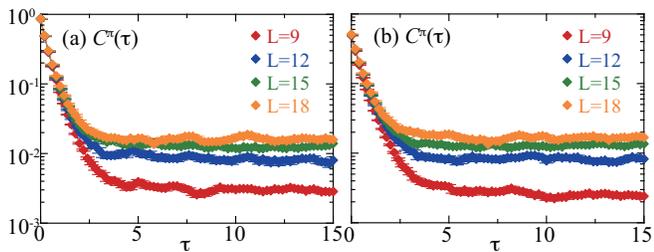}
  \caption{\label{fig:CSO5-Udep} Time-displaced correlation function
    $C^{\pi}(\tau)$ at the DQCP for (a) $U=0$ and (b) $U=0.2$. Here, $T=1/30$.}
\end{figure}

{\it DQCP.}---Contrary to the multicritical point, fermionic excitations are gapped
at the DQCP.  In the framework of Eq.~(\ref{eq:GN}), this implies that
$|\ve{\varphi}|$ remains finite and only its phase fluctuations need to be
considered. Integrating out the fermions \cite{Abanov00,Tanaka05} then yields a 2+1D
non-linear sigma model with a Wess-Zumino-Witten term that accounts for
the phase dynamics of $\ve{\varphi}/|\ve{\varphi}|$. The relevance of terms in
the field theory that break down the SO(5) symmetry to SO(3)$\times$U(1)
at the critical point can again be addressed using $C^{\pi}(\tau)$. The results in
Fig.~\ref{fig:CSO5-Udep} were obtained at different values of $U$. As for the
multicritical point, they suggest an emergent SO(5) symmetry.

{\it Discussion.}---The key result our work is the phase diagram
(Fig.~\ref{fig:phasediagram}) with a fermionic DSM-QSHI-SSC multicritical point as
well as a (bosonic) QSHI-SSC DQCP. Up to the system size accessible in our simulations, we find that
the operator $\hat{\pi}$ that rotates between QSHI and SSC order is a constant of motion of the
low-energy effective Hamiltonian at the critical points. This implies identical anomalous dimensions in the QSHI and SSC
channels, as verified here at the multicritical point and previously at the DQCP
\cite{Liu18}. The substantially different values at the DQCP ($\eta^\alpha
\simeq 0.25$ \cite{Liu18} vs. $\eta^\alpha \simeq 1$) exclude the possibility
that the results at the latter are due to proximity to the multicritical point.  
According to Noether's theorem,  $\hat{\pi}$  is  the  zeroth  component 
of a  conserved   current.  Being  a  conserved  quantity,  it  cannot  acquire  an 
anomalous  dimension.  Such a  criterion  has  been used   to detect  emergent 
SO(4)  symmetry  in Ref.~\cite{Ma19}.

Our findings for the multicritical point give numerical confirmation of
predictions of an emergent SO(5) symmetry based on one-loop RG calculations
\cite{Janssen18,Roy18}. Although we can provide roughly the same quality of
results for the DQCP case, some care has to be taken in interpreting the results
for the latter. Enhanced SO(4) \cite{Zhao19,SatoT17,Ma19} or SO(5) \cite{Nahum15_1,Nahum15}
symmetries have been observed at critical points in various models. In the
SO(4) case, the transition is argued to be of first order
\cite{Zhao19}. Emergent symmetries can occur at first-order transitions due to
fine-tuning.  For instance, the spin-flop transition in an SO(3) symmetric
Heisenberg model has SO(3) symmetry at the UV scale at the transition point. For
weakly first-order transitions \cite{SatoT17,Torres19,Zhao22}, we can understand
emergent symmetries within the RG framework.  In this case, the RG flow becomes
very slow \cite{Nahum15} when approaching the transition.  If the operators that
break an emergent symmetry have a large scaling dimension, they will be
suppressed at intermediate length scales. Therefore, first-order transitions can
be naturally reconciled with emergent symmetries without invoking fine-tuning.

\begin{acknowledgments}
  The authors gratefully acknowledge the Gauss Centre for Supercomputing
  e.V. for funding this project by providing computing time on the GCS
  Supercomputer SUPERMUC-NG at Leibniz Supercomputing Centre.  T.S. acknowledges
  funding from the Deutsche Forschungsgemeinschaft under Grant No.~SA 3986/1-1.
  F.F.A. acknowledges the DFG for funding via W\"urzburg-Dresden Cluster of
  Excellence on Complexity and Topology in Quantum Matter ct.qmat (EXC 2147,
  Project ID 390858490) as well as the SFB1170 on Topological and Correlated
  Electronics at Surfaces and Interfaces.  Y.L. was supported by the National 
  Natural Science Foundation of China under grant No.~11947232.  
  D.H. and W.G. were supported by the National Natural Science Foundation 
  of China under grants No.~12175015 and No.~11734002.
\end{acknowledgments}

%

\clearpage

\section{Supplemental Material}

\section{Free-energy derivative and SSC-QSHI susceptibility ratio}

\begin{figure}[t]
  \centering
  \includegraphics[width=0.475\textwidth]{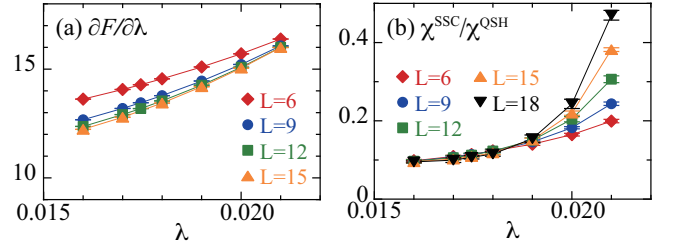}
  \caption{\label{fig:FEdelta} (a) Free-energy derivative and (b) ratio of
    QSHI and SSC susceptibilities $\chi^{\text{SSC}}/\chi^{\text{QSHI}}$ close to
    the multicritical point ($\lambda=0.0174$, $U=0.4$). Here,
    $T=1/30$, representative of the ground state for the parameters considered.}
\end{figure}

In the main text, we discuss a multicritical point with an emergent SO(5)
symmetry at which DSM, QSHI, and SSC phases meet. The scaling behavior observed in
Fig.~\ref{fig:Rchai} is consistent with a continuous phase
transition. This can be substantiated by calculating the free-energy derivative
\begin{equation}
  \frac{\partial F}{\partial \lambda} =  -\frac{1}{L^2}  \left\langle \sum_{\hexagon}
  \left( \sum_{ \langle \langle \ve{i},\ve{j} \rangle \rangle  \in \hexagon
  }\hat{J}_{\bm{i},\bm{j}}\right)^2 \right\rangle\,,
  \label{eq:dfdr-SM}
\end{equation}
which shows no sign of a discontinuity [Fig.~\ref{fig:FEdelta}(a)].

To verify whether the multicritical point has an emergent SO(5) symmetry, we
also measured the ratio of the QSHI and SSC susceptibilities $\chi^{\alpha}$.
These quantities are in general independent but
become locked together and can be combined into a five-component order parameter
if a unifying SO(5) symmetry emerges. In this case, the
ratio $\chi^{\text{SSC}}/\chi^{\text{QSHI}}$ is expected to become scale
invariant, consistent with the data in Fig.~\ref{fig:FEdelta}(b).

\section{Finite-size scaling at the Gross-Neveu and deconfined critical points}

The phase boundaries in Fig.~\ref{fig:phasediagram} are based on results for
the correlation ratios $R^{\text{SSC}}$ and $R^{\text{QSHI}}$. Here, we present
typical results for the DSM-QSHI and DSM-SSC Gross-Neveu transitions as well as
for the QSHI-SSC DQCP.

\begin{figure}[t]
  \centering \includegraphics[width=0.48\textwidth]{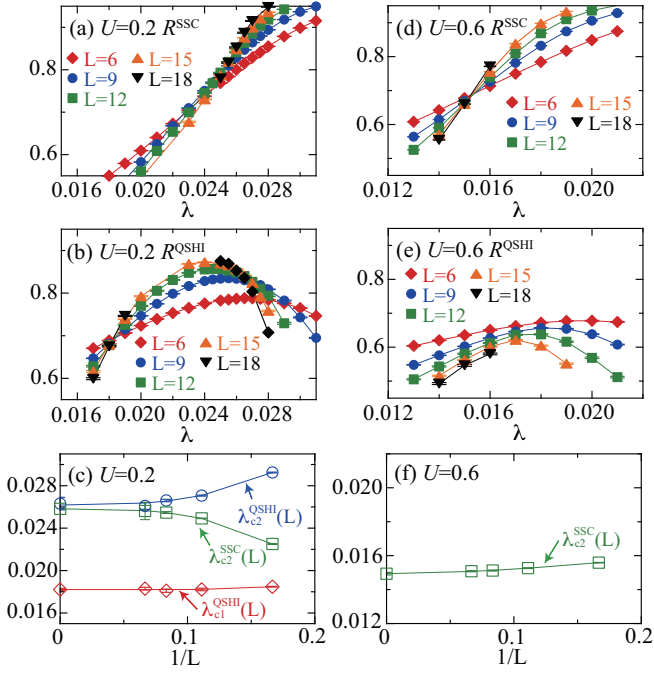}
  \caption{\label{fig:Rchai-SM} Correlation ratios $R^{\text{SSC}}$ [(a),(d)] and
    $R^{\text{QSHI}}$ [(b),(e)] for different system sizes $L$ at fixed $\beta=L$. Extrapolation of the
    crossing points $\lambda^{\alpha}_{c1} (L)$ and $\lambda^{\alpha}_{c2} (L)$ between curves for $L$ and $L+3$ gives
    the critical values reported in Fig.~\ref{fig:phasediagram} [(c),(f)].}
\end{figure}

Figures~\ref{fig:Rchai-SM}(a)--\ref{fig:Rchai-SM}(c) show results for $U=0.2$.
The data for $R^\text{SSC}$ in Fig.~\ref{fig:Rchai-SM}(a) are consistent with a
DSM-SSC transition at $\lambda_{c2}^\text{SSC}=0.0258(1)$, the value obtained by
extrapolating the crossing points to $L\to\infty$ [Fig.~\ref{fig:Rchai-SM}(c)].
Figure~\ref{fig:Rchai-SM}(b) reveals two phase transitions and an intermediate
QSHI phase.  The extrapolation of the two sets of crossing point in
Fig.~\ref{fig:Rchai-SM}(c) yields $\lambda_{c1}^\text{QSHI}=0.0182(2)$ and
$\lambda_{c2}^\text{QSHI}=0.0263(6)$, respectively.  Within error bars,
$\lambda_{c2}^\text{SSC}$ and $\lambda_{c2}^\text{QSHI}$ are identical,
consistent with a direct QSHI-SSC transition.

In Figs.~\ref{fig:Rchai-SM}(d)--\ref{fig:Rchai-SM}(f), we report results at $U=0.6$.
The data for $R^\text{SSC}$ and $R^\text{QSHI}$ indicate a single phase
transition without an intermediate QSHI phase. The extrapolation of the crossing
points [Fig.~\ref{fig:Rchai-SM}(f)] yields $\lambda_{c2}^\text{SSC}=0.0149(1)$ for
the DSM-SSC transition.

\section{Mass terms and the $\pi$ operator}

In the continuum limit, the tight-binding model on the honeycomb lattice reduces
to the Dirac Hamiltonian
\begin{eqnarray}\nonumber
  \hat{H}_0 = -v_\text{F}
  \sum_{\substack{\ve{k},\tau,\tau',\mu,\mu',\sigma}}
  &&\hat{c}^{\dagger}_{\tau,\mu \ve{K} +  \ve{k }, \sigma }
 (   k_x \tau^{x}_{\tau,\tau'}  \mu^{z}_{\mu,\mu'} \\
  && -k_y \tau^y_{\tau,\tau'} \delta_{\mu,\mu'}  )
  \hat{c}^{\phantom\dagger}_{\tau',\mu' \ve{K} +  \ve{k }, \sigma }  \,.
\end{eqnarray}
Here, $\ve{K} = \frac{4}{3} \ve{b}_1 + \frac{4}{3} \ve{b}_2 $ with
$\ve{b}_i $ the reciprocal lattice vectors satisfying
$\ve{b}_i \cdot \ve{a}_j = 2 \pi \delta_{i,j}$.  The Fermi velocity 
$v_\text{F} = \sqrt{3}a t/{2} $ with $a$ the lattice spacing and $t$ the hopping
matrix element.  The indices $\tau$, $\mu$, and $\sigma$ refer to orbital,
valley, and spin, respectively, and take on values $\{-1,1\}$. $\tau^{\alpha}$,
$\mu^{\alpha}$, and $\sigma^{\alpha}$ are Pauli matrices satisfying the
Clifford algebra, $\left\{ \tau^{\alpha} , \tau^{\beta} \right\} = 2
\delta_{\alpha,\beta} $, etc.

In the presence of superconducting mass terms, it is convenient to adopt a Bogoliubov basis
\begin{equation}
  \hat{\Psi}^{\dagger}_{\nu,\mu,\tau,\sigma}(\ve{k}) =  \left\{
    \begin{array}{ll}
      \hat{c}^{\dagger}_{\tau, \mu \ve{K} + \ve{k}, \sigma} &   \text{ if }   \nu = +1 \,, \\
      \hat{c}^{\phantom\dagger}_{\tau, \mu \ve{K} - \ve{k}, \sigma} &   \text{ if }   \nu =  -1\,,
    \end{array}
  \right.
\end{equation}
in which the Hamiltonian takes the form
\begin{eqnarray}
  \hat{H}_0   =     -\frac{v_\text{F}}{2}
  \sum_{\substack{\ve{k},\nu,\nu',\tau,\tau'\\\mu,\mu',\sigma}}
  &&\hat{\Psi}^{\dagger}_{\nu,\tau,\mu,\sigma}  (\ve{k})
     (   k_x \delta_{\nu,\nu'} \tau^{x}_{\tau,\tau'}  \mu^{z}_{\mu,\mu'}
      \\\nonumber
  &&  - k_y \nu^{z}_{\nu,\nu'} \tau^y_{\tau,\tau'} \delta_{\mu,\mu'})
                  \hat{\Psi}^{\phantom\dagger}_{\nu',\tau',\mu', \sigma } (\ve{k})\,.
\end{eqnarray}
We introduced another set of Pauli matrices $\nu^{\alpha}$ acting on the Bogoliubov index.

Let us define mass terms as
\begin{equation}
  \hat{M}   =   \sum_{\ve{k}} \hat{\Psi}^{\dagger}(\ve{k})   M  \hat{\Psi}(\ve{k})\,,
\end{equation}
where $M= M_{\nu,\nu',\tau,\tau',\mu,\mu',\sigma,\sigma'}$ is a $16\times 16 $ matrix accounting for the Bogoliubov, valley, orbital, and spin indices.
In this basis, the three QSHI mass terms read
\begin{eqnarray}
  M_x^\text{QSHI} &=& \nu^z \sigma^x\mu^z\tau^z\,,    \nonumber   \\
  M_y^\text{QSHI} &=& \phantom{\nu^z} \sigma^y\mu^z\tau^z\,,   \nonumber   \\
  M_z^\text{QSHI} &=&  \nu^z \sigma^z\mu^z\tau^z\,.
\end{eqnarray}
The two SC masses, corresponding to the real and imaginary parts of
the SSC order parameter, are
\begin{eqnarray}
  M_1^\text{SSC} &=& \nu^y \sigma^y\mu^x\,,     \nonumber   \\
  M_2^\text{SSC} &=&   \nu^x \sigma^y\mu^x\,.
\end{eqnarray}
By definition, these mass terms anti-commute with $\hat{H}_0$ and mutually anti-commute.

We are now in a position to compute the modes that rotate between QSHI
and SSC order. We focus on
\begin{equation}
  \Pi  =  \frac{i}{2} \left[  M_1^\text{SSC}, M_z^\text{QSHI}   \right]  = - \nu^{x} \sigma^{x}\mu^{y} \tau^{z}.
\end{equation}
The corresponding operator is given by
\begin{equation}
  \hat{\Pi} =   - 2 i \sum_{\substack{\ve{k},\tau,\tau'\\\sigma,\sigma'}}
  \hat{c}^{\dagger}_{\tau, -\ve{K}  - \ve{k}, \sigma}
  \sigma^{x}_{\sigma,\sigma'} \tau^{z}_{\tau,\tau'}
  \hat{c}^{\dagger}_{\tau', \ve{K} +  \ve{k}, \sigma'}      +  \text{H.c.}
\end{equation}
A regularization on the honeycomb lattice takes the form given in
Eq.~(\ref{Eq:pi-mode_main}), $\hat{\pi}=\sum_{\ve{r}}\hat{\pi}_{\ve{r}}$ with
\begin{widetext}
\begin{eqnarray}
  \label{Eq:pi-mode}
  \hat{\pi}_{\ve{r}} =   \hat{\ve{a}}^{\dagger}_{\ve{r}} \sigma^x  \hat{\ve{a}}^{\dagger}_{\ve{r} + \ve{a}_1}
                 +  \hat{\ve{a}}^{\dagger}_{\ve{r}+ \ve{a}_1} \sigma^x  \hat{\ve{a}}^{\dagger}_{\ve{r} + \ve{a}_2}
                 +  \hat{\ve{a}}^{\dagger}_{\ve{r}+ \ve{a}_2} \sigma^x  \hat{\ve{a}}^{\dagger}_{\ve{r} } 
  -    \hat{\ve{b}}^{\dagger}_{\ve{r}} \sigma^x  \hat{\ve{b}}^{\dagger}_{\ve{r} + \ve{a}_1}
         +  \hat{\ve{b}}^{\dagger}_{\ve{r}+ \ve{a}_1} \sigma^x  \hat{\ve{b}}^{\dagger}_{\ve{r} + \ve{a}_2}
         +  \hat{\ve{b}}^{\dagger}_{\ve{r}+ \ve{a}_2} \sigma^x  \hat{\ve{b}}^{\dagger}_{\ve{r} }   + \text{H.c.}
\end{eqnarray}
\end{widetext}
Here,
$ \hat{\ve{a}}^{\dagger}_{\ve{r}} = (\hat{a}^{\dagger}_{\ve{r},\uparrow},
  \hat{a}^{\dagger}_{\ve{r},\downarrow} ) $ and
$ \hat{\ve{b}}^{\dagger}_{\ve{r}} = (\hat{b}^{\dagger}_{\ve{r},\uparrow},
  \hat{b}^{\dagger}_{\ve{r},\downarrow}) $. The operators
$ \hat{a}^{\dagger}_{\ve{r},\sigma} $, and $ \hat{b}^{\dagger}_{\ve{r},\sigma} $
create fermions in the $A$ and $B$ orbitals of unit cell $\ve{r}$, respectively;
$\ve{a}_1 = a (1,0) $ and $\ve{a}_2 = a \left( 1/2, \sqrt{3}/{2} \right) $ are the basis vectors.

We now discuss the symmetry properties of $\hat{\Pi}$.

\paragraph{Time reversal:}

Since the SSC and QSHI mass terms are even under time reversal, $\hat{\Pi}$ has
to be odd. This becomes apparent from Eq.~(\ref{Eq:pi-mode})  since  $\sigma_x$
maps onto  $- \sigma_x$ under time reversal.

\paragraph{Inversion:}

Inversion around the  center  of  a  hexagon leaves the SSC order  parameter
invariant.  On the  other hand, the QSHI order parameters are odd  under this
operation.  Hence, $\hat{\Pi}$ is odd  under  inversion. Accordingly,
in  Eq.~(\ref{Eq:pi-mode}), interchanging $\hat{\vec{a}}$ and $\hat{\vec{b}}$ flips the sign
of $\hat{\pi}_{\ve{r}}$.

\paragraph{Spin rotations:}

To understand spin rotations, we note that defining
\begin{equation}
  U(\ve{e},\theta) = e^{ -i \theta \ve{e} \cdot \ve{\sigma}/2 } 
\end{equation}
  we have
\begin{eqnarray}\label{eq:spinrotations}
  U^{\dagger} (\ve{e},\theta)   \ve{\sigma}   \sigma_y \left[ U^{\dagger} (\ve{e},\theta)  \right]^{T}
  & & =  U^{\dagger} (\ve{e},\theta )   \ve{\sigma}   U (\ve{e},\theta )   \sigma_y   \nonumber  \\
  & & = R(\ve{e},\theta) \ve{\sigma}   \sigma_y
\end{eqnarray}
with $R(\ve{e},\theta) $ an SO(3) rotation around axis $\ve{e}$ with angle
$\theta$.  Using Eq.~(\ref{eq:spinrotations}), it can be verified that the SSC
mass term is a spin singlet and hence invariant under the transformation:
$ \hat{c}_{\tau,\mu \ve{K} + \ve{k }, \sigma } \rightarrow \sum_{\sigma'}
U(\ve{e},\theta)_{\sigma,\sigma'} \hat{c}_{\tau,\mu \ve{K} + \ve{k }, \sigma' }
$.  On the other hand, the QSHI mass terms transform as a vector,
$\hat{\ve{M}}^\text{QSHI} \rightarrow R(\ve{e},\theta) \hat{\ve{M}}^\text{QSHI} $.  Writing
$\sigma^x = \ve{e}^z \cdot \ve{\sigma} i \sigma^{y} $, it becomes apparent that
$\hat{\pi}$ transforms as the $z$-component of the QSHI masses.

\section{$\tau$-independence of $C^{\pi}(\tau)$ and emergent symmetry}

Consider the energy basis representation,
$ \hat{H} | n, \lambda_n \rangle = E_n | n, \lambda_n \rangle $.  Here, we allow
for degeneracies $\lambda_n$ of the energy eigenstates.
In this basis, time-displaced correlation functions of a hermitian operator read
\begin{equation}
  \label{Eq:otau}
  \langle \hat{O} (\tau)    \hat{O}   \rangle  =    \frac{\sum_{n,\lambda_n, m, \lambda_m}     e^{-\beta  E_n  }  e^{\tau  \left[
        E_n  - E_m  \right] }  |  \langle n, \lambda_n  |  \hat{O} | m, \lambda_m \rangle|^2 }{ \sum_{n,\lambda_n}  e^{- \beta E_n  }}\,.
\end{equation}
We will assume that for $\tau > \tau_1$, with $ 0 < \tau_1 < \beta/2 $, the
correlation function is constant up to exponentially small
corrections and set $E_0 = 0$. Then, $1/\Lambda = \text{min}(\beta - \tau_1,
\tau_1) $ defines an inverse energy scale. We can define a projection onto a low-energy Hilbert space,
\begin{equation}
  \hat{P}  =  \sum_{n, \lambda_n, E_n <  \Lambda}    |  n , \lambda_n \rangle  \langle  n, \lambda_n |\,,
\end{equation}
and a low-energy effective Hamiltonian,
\begin{equation}
  \hat{H}_\text{eff}    =  \hat{P} \hat{H}  \hat{P}\,.
\end{equation}
Following our initial assumption,
\begin{equation}
  \langle \hat{O} (\tau)    \hat{O}   \rangle_\text{eff}   =   \frac{\text{Tr}  \left[  \hat{P} e^{-(\beta - \tau) \hat{H}} \hat{O} \hat{P} e^{-\tau{H}} \hat{O} \right] }{
    \text{Tr}  \left[  \hat{P} e^{- \beta \hat{H}} \right] }
\end{equation}
does not depend on $\tau$.

As a consequence of Eq.~(\ref{Eq:otau}), and for energy eigenstates in the
low-energy Hilbert space, $\hat{O} | m,\lambda_m \rangle $ has to be included in
the eigenspace ${\cal H}_{E_m} $ of the $E_m$ eigenvalue:
$\left\{ | m ,\lambda_m \rangle , \lambda_m = 1\, \dots, \dim {\cal H}_{E_m}
\right\} $.  In other words,
\begin{equation}
  \hat{O} | m, \lambda_m \rangle    =   \sum_{\lambda_{m'}} O_{\lambda_m,\lambda_{m'}}  | m, \lambda_{m'} \rangle.  
\end{equation} 
We can now perform a unitary transformation in each eigenspace to
construct the basis $\tilde{ | m, \lambda_m \rangle} $ satisfying
\begin{eqnarray}
  \hat{H}_\text{eff}\tilde{ | m, \lambda_m \rangle}   &  = &   E_m  \tilde{ | m, \lambda_m \rangle}\,, \nonumber  \\
  \hat{O}\tilde{ | m, \lambda_m \rangle}   & =  &  O_{m,\lambda_m} \tilde{ | m, \lambda_m \rangle}. 
\end{eqnarray} 
Hence,
\begin{equation}\label{eq:commutation}
  \left[  \hat{H}_\text{eff} ,  \hat{O} \right]   = 0. 
\end{equation}
 
Equation~(\ref{eq:commutation}) implies that 
$\langle \hat{O} (\tau) \hat{O} \rangle_\text{eff} = \langle \hat{O} \hat{O}
\rangle_\text{eff} $ is $\tau$-independent.  For our specific case, the operator
$\hat{O} = \hat{\pi} $ transforms as the $z$-component of a vector under spin
rotations.  Since $ \hat{H}_\text{eff}$ has the same symmetries as $\hat{H}$, and
assuming that the ground state is a spin singlet, we have $ \hat{\pi} |0 \rangle = 0 $ and
$\lim_{\beta \rightarrow \infty} \langle \hat{{\pi}} \cdot \hat{{\pi}} \rangle =
|| {\hat{\pi}} |0 \rangle ||^{2} = 0 $.  In contrast, at finite temperatures,
there is no symmetry argument for
$\langle \hat{{\pi}} \cdot \hat{{\pi}} \rangle$ to vanish.

\section{Bubble approximation of $C^{\pi}(\tau)$}

In the main text, we reported results for the imaginary-time correlation
function of $\hat{\pi}$ [Eq.~(\ref{Eq:pi-mode})],
\begin{eqnarray}
  \label{Eq:pi-mode-c}
  C^{\pi}(\tau) &=& \frac{1}{L^2}\sum_{\ve{i},\ve{j}} \langle
                    \hat{\pi}_{\ve{i}}(\tau) \hat{\pi}_{\ve{j}}(0)\rangle\\
               &=& \langle \hat{\pi}_{\ve{q}}(\tau)   \hat{\pi}_{-\ve{q}}(0)\rangle_{\ve{q}=\ve{0}}\nonumber
\end{eqnarray}
with
\begin{widetext}
\begin{eqnarray}
  \label{Eq:pi-mode-k}
  \hat{\pi}_{\ve{q}} (\tau) = \frac{1}{L}\sum_{l}(-1)^{l}\sum_{\ve{k}} 
  &&\Big[ \Big( e^{-i \left( \ve{k}-\ve{q} \right) \cdot \ve{a}_1}
  +e^{i  \left(\ve{k} \cdot \ve{a}_1-\left( \ve{k}- \ve{q}\right)\cdot \ve{a}_2\right) }
  +e^{i \ve{k}\cdot \ve{a}_2} \Big)
                         \hat{\ve{c}}^{\dagger}_{\ve{k},l}(\tau)  \sigma^x  \hat{\ve{c}}^{\dagger}_{-\ve{k}+\ve{q},l}(\tau)\nonumber  \\
    &&+\Big( e^{i \left( \ve{k}+\ve{q} \right) \cdot \ve{a}_1}
  +e^{-i  \left(\ve{k} \cdot \ve{a}_1-\left( \ve{k}+ \ve{q}\right)\cdot \ve{a}_2\right) }
  +e^{-i \ve{k}\cdot \ve{a}_2} \Big)
                         \hat{\ve{c}}^{\phantom\dagger}_{-\ve{k}-\ve{q},l}(\tau)  \sigma^x  \hat{\ve{c}}^{\phantom\dagger}_{\ve{k},l}(\tau)\Big]\,.
\end{eqnarray}
\end{widetext}
Here, $ \hat{\ve{c}}^{\dagger}_{\ve{k},l} =
(\hat{c}^{\dagger}_{\ve{k},l,\uparrow},
  \hat{c}^{\dagger}_{\ve{k},l,\downarrow} ) $, $l\in\{A, B\}$ is an orbital
index and $(-1)^{l} = 1 $ for $l=A$ and $-1$ for $l=B$.

Using Eqs.~(\ref{Eq:pi-mode-c}) and (\ref{Eq:pi-mode-k}), $C^{\pi}(\tau)$ is given by
\begin{widetext}
\begin{eqnarray}
  \label{Eq:pi-mode-ck}
  C^{\pi}(\tau)=\frac{1}{L^2}\sum_{l,l'}(-1)^{l+l'} \sum_{\ve{k},\ve{k'}}v_{\ve{k}} v_{\ve{k'}}^{*}
                \Big[ \langle \hat{\ve{c}}^{\dagger}_{\ve{k},l}(\tau)  \sigma^x
  \hat{\ve{c}}^{\dagger}_{\ve{-k},l}(\tau)\hat{\ve{c}}^{\phantom\dagger}_{\ve{-k'},l'}(0)
  \sigma^x  \hat{\ve{c}}^{\phantom\dagger}_{\ve{k'},l'}(0)\rangle 
 +
                  \langle 
                  \hat{\ve{c}}^{\phantom\dagger}_{\ve{-k'},l}(\tau)
  \sigma^x  \hat{\ve{c}}^{\phantom\dagger}_{\ve{k'},l}(\tau)
  \hat{\ve{c}}^{\dagger}_{\ve{k},l'}(0)  \sigma^x
  \hat{\ve{c}}^{\dagger}_{\ve{-k},l'}(0) \rangle\Big]\nonumber  \\
\end{eqnarray}
\end{widetext}
where
$v_{\ve{k}}=e^{-i \ve{k} \cdot \ve{a}_1}+e^{i \ve{k} \cdot \left(
    \ve{a}_1-\ve{a}_2\right) }+e^{i \ve{k} \cdot \ve{a}_2}$.
In the main text, we included results for $C^{\pi}(\tau)$ that neglect vertex corrections, as
calculated from a convolution of the single-particle Green function and
\begin{widetext}
\begin{eqnarray}
  \label{Eq:pi-mode-ckbubble}
  C^{0,\pi}(\tau)&=& \frac{1}{L^2} \sum_{l,l'}(-1)^{l+l'}
                \\
               &&
                \times     \Big[
                     \sum_{\ve{k}} v_{\ve{k}} v_{\ve{k}}^{*}
                    \Big(
                    \langle
                    \hat{\ve{c}}^{\dagger}_{\ve{k},l,\sigma_1}(\tau)\hat{\ve{c}}^{\phantom\dagger}_{\ve{k},l'\sigma_4}(0)
                    \rangle
                    \langle
                    \hat{\ve{c}}^{\dagger}_{\ve{-k},l,\sigma_2}(\tau)\hat{\ve{c}}^{\phantom\dagger}_{\ve{-k},l',\sigma_3}(0)\rangle
                +
                    \langle
                    \hat{\ve{c}}^{\phantom\dagger}_{\ve{k},l,\sigma_1}(\tau)\hat{\ve{c}}^{\dagger}_{\ve{k},l'\sigma_4}(0)
                    \rangle
                    \langle
                    \hat{\ve{c}}^{\phantom\dagger}_{\ve{-k},l,\sigma_2}(\tau)\hat{\ve{c}}^{\dagger}_{\ve{-k},l',\sigma_3}(0)\rangle
                    \Big)
                     \nonumber  \\
  &&\quad-\sum_{\ve{k}} v_{\ve{k}} v_{\ve{-k}}^{*}
                \Big(
                    \langle
                    \hat{\ve{c}}^{\dagger}_{\ve{k},l,\sigma_1}(\tau)\hat{\ve{c}}^{\phantom\dagger}_{\ve{k},l'\sigma_3}(0)
                    \rangle
                    \langle
                    \hat{\ve{c}}^{\dagger}_{\ve{-k},l,\sigma_2}(\tau)\hat{\ve{c}}^{\phantom\dagger}_{\ve{-k},l',\sigma_4}(0)\rangle
                +
                    \langle
                    \hat{\ve{c}}^{\phantom\dagger}_{\ve{k},l,\sigma_1}(\tau)\hat{\ve{c}}^{\dagger}_{\ve{k},l'\sigma_3}(0)
                    \rangle
                    \langle
                    \hat{\ve{c}}^{\phantom\dagger}_{\ve{-k},l,\sigma_2}(\tau)\hat{\ve{c}}^{\dagger}_{\ve{-k},l',\sigma_4}(0)\rangle
                    \Big) \Big]\nonumber\,,
\end{eqnarray}
\end{widetext}
where
$(\sigma_1,\sigma_2,\sigma_3,\sigma_4)$ takes on the values $(\uparrow,\downarrow,\downarrow,\uparrow)$,
${(\uparrow,\downarrow,\uparrow,\downarrow)}$,
$(\downarrow,\uparrow,\uparrow,\downarrow)$, and $(\downarrow,\uparrow,\downarrow,\uparrow)$.

\end{document}